# Two Dimensional Ising Superconductivity in Gated MoS$_2$

J. M. Lu[1], O. Zeliuk[1], I. Leermakers[2], Noah F. Q. Yuan[3], U. Zeitler[2], K. T. Law[3] and J. T. Ye[1*]

[1] *Device Physics of Complex Materials, Zernike Institute for Advanced Materials,*

*Nijenborgh 4, 9747 AG, Groningen, The Netherlands*

[2] *High Field Magnet Laboratory (HFML-EMFL), Radboud University,*

*Toernooiveld 7, 6525 ED Nijmegen, The Netherlands*

[3] *Department of Physics, Hong Kong University of Science and Technology,*

*Clear Water Bay, Hong Kong, China*

**The Zeeman effect, which is usually considered to be detrimental to superconductivity, can surprisingly protect the superconducting states created by gating a layered transition metal dichalcogenide. This effective Zeeman field, which is originated from intrinsic spin orbit coupling induced by breaking in-plane inversion symmetry, can reach nearly a hundred Tesla in magnitude. It strongly pins the spin orientation of the electrons to the out-of-plane directions and protects the superconductivity from being destroyed by an in-plane external magnetic field. In magnetotransport experiments of ionic-gate MoS$_2$ transistors, where gating prepares individual superconducting state with different carrier doping, we indeed observe a spin-protected superconductivity by measuring an in-plane critical field $B_{c2}$ far beyond the Pauli paramagnetic limit. The gating-enhanced $B_{c2}$ is more than an order of magnitude larger compared to the bulk superconducting phases where the effective Zeeman field is weakened by interlayer coupling. Our study gives the first experimental evidence of an Ising superconductor, in which spins of the pairing electrons are strongly pinned by an effective Zeeman field.**

[*] email address: j.ye@rug.nl

In conventional superconductors, applying a sufficiently high magnetic field above the upper critical field $B_{c2}$ is a direct way to destroy superconductivity by breaking Cooper pairs via the coexisting orbital and Pauli paramagnetic mechanisms. The orbital contribution originates from the coupling between the magnetic field and the electron momentum, whereas the paramagnetic contribution is caused by spin alignment in Cooper pairs by an external magnetic field. When the orbital effect is weakened or eliminated, either by having large electron mass (1) or reducing dimensionality (2), $B_{c2}$ is solely determined by the interaction between the magnetic field and the spin degree of freedom of Cooper pairs. In superconductors where Cooper pairs are formed by electrons with opposite spin, energy is gained from aligning the electron spins by the external magnetic field, and, therefore, $B_{c2}$ cannot exceed the Clogston-Chandraskhar (3,4) or the Pauli paramagnetic limit (in unit of T) $B_p \approx 1.86\, T_c(0)$. Here, $T_c(0)$ is zero-field superconducting critical temperature (in unit of K) characterizing the binding energy of a Cooper pair, which competes with the Zeeman splitting energy.

However, in some superconductors the Pauli limit can be bypassed in different ways. For example, forming Fulde–Ferrell–Larkin–Ovchinnikov (FFLO) states with inhomogeneous pairing densities favors the presence of a magnetic field even above $B_p$ (5). In spin triplet superconductors, the parallel aligned spin configuration in Cooper pairs is not affected by Pauli paramagentism. Therefore, the $B_{c2}$, which is only determined by orbital depairing mechanism, can easily exceed the $B_p$ in many cases (6–8). Also spin-orbit interaction has been shown to align spins to overcome the Pauli limit. Rashba spin-orbit coupling (SOC) in non-centrosymmetric superconductors will lock the spin to the in-plane direction, which can greatly enhance the out-of-plane $B_{c2}$ (9). However, for in-plane magnetic field, $B_{c2}$ still suffers from weakened paramagnetism, and therefore, the in-plane $B_{c2}$ protected by Rashba SOC can only be moderately enhanced to $\sqrt{2}B_p$ (10). Alternatively, electron spins can be randomized by spin orbit scattering (SOS), which consequently weakens the effect of spin paramagnetism (11–15) and enhances $B_{c2}$.

Recently, it has been shown that superconductivity in thin flakes of $MoS_2$ can be induced electrostatically using the electric field effect mediated by moving ions in a voltage biased ionic liquid placed on top of the sample (as detailed in section 1 of the Supporting Online Materials, SOM). Negative carriers (electrons) are induced by accumulating cations above the outermost layer of $MoS_2$ flake forming a capacitor of ~1 nm thick (16–21). The potential gradient at the surface creates a planar homogenous electronic system with an inhomogeneous vertical doping profile where conducting electrons are predominately doped into a few outermost layers forming superconducting states near the *K* and *K'* valleys of conduction band as schematically shown in Fig. 1A. The in-plane inversion symmetry breaking in a $MoS_2$ monolayer can induce SOC manifested as a Zeeman–like effective magnetic field $\boldsymbol{B}_{eff}$ (~100 T) being oppositely applied at *K* and *K'* points of the Brillouin Zone (22). Since electrons of opposite momentum experiences opposite $\boldsymbol{B}_{eff}$, this SOC is then compatible with Cooper pairs also residing at *K* and *K'* points (23). Therefore, spins of electrons in the Cooper pairs are polarized by this large out-of-plane Zeeman field which is able to protect their orientation from being realigned by an in-plane magnetic field. Consequently, gate-induced superconducting $MoS_2$ should exhibit a large in-plane $B_{c2}$. This unique alternating spin configuration also provides the essential ingredient for establishing an Ising superconductor where spins of electrons in the Cooper pairs are strongly pined by an effective Zeeman field in Ising-like fashion.

Nevertheless, due to the alternating stacking order in 2*H*-type TMD single crystals (Fig. 1B), electrons with the same momentum experience $\boldsymbol{B}_{eff}$ with opposite signs for adjacent layers, which weaken the effect of SOC by cancelling out $\boldsymbol{B}_{eff}$ mutually in bulk crystal (Fig. 1C). Given our multi-layered system, however, field-effect doping can confine carriers strongly at the outermost layer reaching 2D carrier density $n_{2D}$ up to ~$10^{14}$ cm$^{-2}$ (21,24). As shown in Fig. 1D (left axis), *Ab initio* calculation of our devices showed that $n_{2D}$ of individual layer decays exponentially from the channel surface. Consequently, $n_{2D}$ of the second layer reduced by almost 90% compared to the first one (25). From the established phase diagram (21), if superconductivity is induced close to the quantum critical point (QCP, $n_{2D} \sim 6\times10^{13}$ cm$^{-2}$), the second layer from the outmost is not even metallic because metallic transport can be observed only when $n_{2D} > 8\times10^{12}$ cm$^{-2}$. Therefore, the outermost layer is well isolated by gating from the rest layers mimicking a freestanding monolayer (26). As shown in Fig. 1D (right axis), in this work, superconducting states with different doping can be obtained by selecting proper gating (21), which exhibit different $T_c$ (at $B = 0$) in the temperature dependence of sheet resistivity $R_s$ (Fig. 1E). A superconducting state ($T_c(0) = 2.37$ K) right at the onset of superconductivity (close to QCP) could be induced without suffering from inhomogeneity usually found at low doping concentration (Fig. 1E, red curve). Consistently, this well-behaved state also exhibits a high mobility ~700 cm$^2$/Vs (measured at $T = 15$ K) before $R_s$ reaching zero resistance.

It is worth noting that ARPES measurements (26,27) and *ab initio* calculation (24,28) both showed that electron doping starts near the *K* points of the conduction band. At sufficiently high doping, the raised Fermi level $E_F$ can touch the local band minima at *Q* points near the middle of *K–Γ* lines as well. The filling of *Q* pockets are also influenced by lowering of band minima at *Q* points as a result of doping-induced modification of band structure (24,28). This means the simplest superconducting states in MoS$_2$ dominated by Cooper pairs at *K* and *K'* points should be prepared by minimized doping.

The charge distribution of our gated system implies that the superconducting state thus formed should exhibit purely 2D nature. To prove this dimensionality, we have characterized sample D24 (with $T_c(0) = 7.38$ K) by a series of measurements. As shown in the temperature dependence of $R_s$ under out-of- and in-plane magnetic field (Fig. 2A and B, respectively), the superconducting state is highly anisotropic. $R_s$ measured as a function of the angle $\theta$ between the magnetic field and the MoS$_2$ plane is shown in Fig. 2C, where $B_{c2}$ was obtained at $T = 6.99$ K for different angle $\theta$ (Fig. 2D). The angular dependence of $B_{c2}$ is fitted by the 2D Tinkham formula (red curve) (29) and a 3D anisotropic Ginzburg-Landau model (blue curve) (2). For $\theta > \pm 1°$, our data seems to be consistent with both models. However, for $\theta < \pm 1°$, as shown in the insert of Fig.2D, the cusp-shaped dependence can only be explained by a 2D model. These measurements unambiguously show that superconducting states are 2D in nature similar to the cases found at LaAlO$_3$/SrTiO$_3$ interface (30) and ion-gated SrTiO$_3$ surface (31). From the *V–I* dependence at different temperatures close to $T_c(0)$ (Fig. 2E), we determined that the Berezinskii-Kosterlitz-Thouless (BKT) temperature $T_{BKT} = 6.3$ K for our 2D system (Fig. 2F). Additional *V-I* characteristics in a magnetic field (Fig. S3, SOM) exhibt a similar critical behavior as the zero-field data, where $T_{BKT}$ is effectively reduced by increasing magnetic field.

From Fig. 2B, given $T_c(0) = 7.38$ K (BCS Pauli limit $B_P = 13.7$ T), it is evident that a moderate in-plane $B$ field of up to 12 T shows little effect on changing the superconducting transition temperature thus $B_{c2}$ of the system must be far above $B_P$. To confirm this speculation, we have performed a high field measurement up to 37 T (see Sec. 2 of SOM for measurement details) on sample D1 after observing a steep increase of $B_{c2}$ near $T_c(0) = 5.5$ K (green dots, Fig. 3C). By controlling the gating strength, superconducting states with $T_c(0) = 2.37$ K and 7.64 K were induced in sample D1. For $T_c(0) = 2.37$ K, we obtained $B_{c2}$ as the magnetic field required to reach $R_N/2$ when $R_s$ at different temperatures are measured as a function of in-plane magnetic field (Fig. 3A). The temperature dependence of $B_{c2}$, shown as red dots in Fig. 3C, is above 20 T at 1.46 K, which is already more than 4 times higher than the $B_P$. Also data from the second gating ($T_c(0) = 7.64$ K, Fig. 3B) shows only weak reduction of $T_c$ of ~1 K at even the highest magnetic field 32.5 T (~ $2.3 \times B_p$) when temperature dependence of $R_s$ was measured at a fixed magnetic field.

The temperature dependence of in-plane $B_{c2}$ for samples D1 in three different states are summarized in Fig. 3C accompanied by theoretical fitting using a phenomenological Ginzburg-Landau (GL) theory in 2D limit (2) and the microscopic Klemm-Luther-Beasley (KLB) theory (12,15,32). The zero temperature in-plane $B_{c2}$ extrapolated for all three superconducting states are far beyond $B_p$. As expected, the zero-temperature $B_{c2}$ predicted by 2D GL theory, without taking spin into account, is larger than that estimated by the KLB theory, which considers both the limiting effect from spin paramagnetism and the enhancing effect by the spin-orbit scattering (SOS) from disorder. To fit the data using KLB theory (dashed curves in Fig. 3C), the interlayer coupling has to be set to zero. This strongly suggests that the induce superconductivity is purely 2D in nature, which is consistent with theoretical calculations (21,25) and ARPES measurements (26,27).

Although the fitted curves can well match our data and scattering time, $\tau_{SO} \sim 40$–50 fs has been observed in organic molecule intercalated $TaS_2$ (37–39), $(LaSe)_{1.14}(NbSe_2)$ (40,41), and organic superconductors $\kappa\text{-}(ET)_4Hg_{2.89}Br_8$ (42) as well, fitting by KLB theory requires very short SOS time to be ~ 24 fs (Fig. S5), which is even less than the total scattering time of 185 fs estimated from resistivity measurement at 15 K (Table S2). This $\tau_{SO}$ is much shorter than the estimation of ~ ns calculated for MoS$_2$ at the carrier density range accessed by this work (43). It is also orders of magnitude shorter than that found in gold film (44–47), where SOS is expected to be much larger in the much heavier gold atoms (Table S2 in SOM).

For comparison, the temperature dependence of $B_{c2}$ of bulk superconducting MoS$_2$ intercalated by alkali metals (48) are also shown in Fig. 3C. For the bulk phase, it is evident that $B_{c2}$ near $T_c$ (0) shows a rather linear instead of square root dependence on temperature, which is consistent with the property of 3D superconductors. The following slight upturn of $B_{c2}$ toward lower temperature away from $T_c$ (0) is the evidence of crossover from 3D to 2D superconducting states (12,32,38–40) due to the layered nature of bulk crystal. Conversely, in all these bulk phases, the measured $B_{c2}$ values are all much smaller or comparable (by doping Cs) with $B_p$ (48).

As a further comparison between bulk and gate-induced superconducting states, the in-plane $B_{c2}$ normalized by $B_p$ is plotted as a function of the reduced temperature $T/T_c$ in Fig. 3D. The dashed line in Fig.3D indicates $B_p$. Clearly, all normalized $B_{c2}$ of bulk superconducting phases fall within the shaded area bounded by the Pauli limit. As discussed above, the large SOS rate used in KLB theory, which should be also present in bulk samples to enhance $B_{c2}$, is not consistent with observed small in-plane $B_{c2}$ for bulk phases, which is clearly limited by $B_p$. Therefore, SOS is highly unlikely to be the origin of the enhancement of $B_{c2}$. In contrast, all gate-induced phases (from samples D1 and 24) are far above both $B_p$ (dashed line) and bulk-phase $B_{c2}$. Interestingly, the D1 with $T_c$ (0) = 2.37 K exhibits the largest enhancement, which is separated from the other gate-induced states.

Excluding SOS as the principal mechanism for the strong enhancement of the in-plane $B_{c2}$ and taking into account of the recent developments in understanding the band structures of monolayer MoS$_2$ (49,50), we propose that this $B_{c2}$ enhancement is mainly due to the intrinsic spin-orbit coupling in MoS$_2$. Near the $K$ points of Brillouin zone (Fig. 1A) and in the basis of spin up and down electrons $[\psi_{k\uparrow}, \psi_{k\downarrow}]$, the normal state Hamiltonian of monolayer MoS$_2$ in the presence of external field can be described by (23)

$$H(\mathbf{k} + \epsilon \mathbf{K}) = \varepsilon_k + \epsilon \beta_{so} \sigma_z + \alpha_R \mathbf{g}_F \cdot \boldsymbol{\sigma} + \mathbf{b} \cdot \boldsymbol{\sigma}$$

Here, $\varepsilon_k = \frac{k^2}{2m} - \mu$ denotes the kinetic energy with chemical potential $\mu$, $m$ is the effective mass of the electrons, $\boldsymbol{\sigma} = (\sigma_x, \sigma_y, \sigma_z)$ are the Pauli matrices, $\boldsymbol{g}_F = (k_y, -k_x, 0)$ denotes the Rashba vector (always lying in-plane), $\alpha_R$ and $\beta_{SO}$ are the strength of Rashba and intrinsic SOC, respectively, $\epsilon = \pm 1$ is the valley index which equals to 1 at the K valley and –1 at the K' valley and $\boldsymbol{b} = \mu_B \boldsymbol{B}$ is the external Zeeman field. The intrinsic SOC term $\epsilon \beta_{SO} \sigma_z$, due to in-plane inversion symmetry breaking induces an effective magnetic field pointing out-of-plane (z direction), which has opposite signs at opposite valleys (shown as green arrows in Fig. 1A). It is worth noting that this Zeeman-like effective magnetic field $\boldsymbol{B}_{\text{eff}} = \epsilon \beta_{SO} \boldsymbol{z} / g \mu_B$ (g is gyromagnetic ratio) will only appear in our multi-layered system after applying a large electric field which isolates the outermost layers from the rest (21,51), thus mimicking a monolayer system. As we will show below, this $\boldsymbol{B}_{\text{eff}}$, equivalent to ~ 100 T, strongly pins the electron spin into out-of-plane direction and protects the electron spins of Cooper pairs from being tilted to the in-plane direction by an external magnetic field, therefore, enhancing the in-plane $B_{c2}$ significantly. On the other hand, large electric field generated by gating reaches ~50 MV/cm (21), which also causes out-of-plane inversion symmetry breaking. Hence, gating also induces a coexisting Rashba–type effective magnetic field $\boldsymbol{B}_{\text{Ra}} = \alpha_R \boldsymbol{g}_F / g \mu_B$ in our system.

When Cooper pairs are influenced by both $\boldsymbol{B}_{\text{eff}}$ and $\boldsymbol{B}_{\text{Ra}}$, the total energy in a magnetic field is schematically shown in Fig. 4A-D. If the electron spin aligned by $\boldsymbol{B}_{\text{eff}}$ ($\boldsymbol{B}_{\text{Ra}}$) stays parallel to the external magnetic field $\boldsymbol{B}_{\text{ex}}$, (Fig. 4A (C)), the system gains energy through coupling between spin and external fields as $\mu_B \cdot B_{ex}$. Therefore, $B_{c2}$ is limited by $B_p$ (Fig. 4A), or can reach $\sqrt{2} B_p$ (Fig. 4C) when coupling is reduced in Rashba-type spin configuration (10), respectively. When $\boldsymbol{B}_{\text{eff}}$ and $\boldsymbol{B}_{\text{Ra}}$ are perpendicular to $\boldsymbol{B}_{\text{ex}}$ as respectively shown in Fig. 4B and D, spin aligned by both effective fields is orthogonal to $\boldsymbol{B}_{\text{ex}}$. Hence, the coupling between spin and $\boldsymbol{B}_{\text{ex}}$ are minimized and $B_{c2}$ can easily bypass $B_p$ in these two cases.

To theoretically describe our system subjected to an in-plane external magnetic field (combining the cases shown in Fig. 4B and C), we introduce the pairing potential terms $\Delta \psi_{k\uparrow} \psi_{-k\downarrow} + h.c.$ into $H(\boldsymbol{k})$ and solve the self-consistent mean field gap equation (Sec. 6 of SOM). The in-plane $B_{c2}$ for a sample with a given $T_c$ can then be determined by including the intrinsic SOC term $\beta_{SO}$ and the Rashba energy $\alpha_R k_F$, where $k_F$ is the Fermi momentum.

For the most extended data from sample D1 ($T_c(0) = 2.37$ K), the relationship between $B_{c2}/B_p$ and the reduced temperature $T/T_c$, shown in Fig. 4E can be perfectly fitted with $\beta_{SO} = 6.2$ meV and $\alpha_R k_F = 0.88$ meV. The value obtained for $\beta_{SO}$ corresponds to an out-of-plane field of ~114 T, which is comparable with the value expected from *ab initio* calculation at the *K* point (3 meV) (22). In fact, to reach superconducting states (21), $E_F$ has to be increased far above the conduction band edge at the *K* points, where a larger $\beta_{SO}$ is expected at higher $E_F$ (52). Meanwhile, the Rashba energy obtained can be regarded as an upper bound of estimation because present model does not include impurity scattering, which can cause the same effect of reducing $B_{c2}$ (36). Despite the simplicity, the in-plane $B_{c2}$ (D1, red curve in Fig. 1E) can be well explained by aforementioned Hamiltonian introduced with little tuning degrees of freedom.

As predicted the model presented, the scale of $B_{c2}$ enhancement is determined by a destructive interplay between intrinsic $\beta_{SO}$ and $\alpha_R k_F$. Reaching higher $T_c(0)$ requires stronger doping under higher electric fields with concomitant increase of $\boldsymbol{B}_{Ra}$. As a result of this competition, the in-plane $B_{c2}$ protection should be weakened with the increase of $T_c(0)$. To support this argument, we choose two other superconducting states that showed consecutive higher $T_c(0)$ (from D1 and D24). By assuming identical $\beta_{SO}$ (6.2 meV), $B_{c2}$ from D1 with $T_c(0) = 5.5$ K and D24 with $T_c(0) = 7.38$ K can be well fitted using $\alpha_R k_F = 1.94$ and 3.02 meV, respectively, which is consistent with the expected increase of $\alpha_R k_F$ together with $T_c(0)$. As shown in Fig. 4E, the increase of the Rashba energy well corroborates our physical picture: the spin-protection is originated from intrinsic $\beta_{SO}$ and influenced by an interplay between $\boldsymbol{B}_{eff}$ and $\boldsymbol{B}_{Ra}$. Accessing even higher $T_c(0)$ requires a further increase of $E_F$, which will eventually occupy the *Q* pocket in the conduction band (24,28,50). Although an even larger $\beta_{SO}$ ~ 80 meV (53) existing at the *Q* point will not alter the aforementioned physical picture obtained at the *K* point, superconducting states could be complicated by having more than one gap. Further doping would also make the adjacent second layer from the top more metallic causing enhanced tunneling between layers, which is beyond the description of our simple two-band model.

It is important to note that aforementioned pining of spin will be weakened by interlayer coupling in bulk phases because $\boldsymbol{B}_{\text{eff}}$ reverses direction in alternating layers in 2$H$-type crystal structure (Fig. 1B and C). After entering bulk superconductivity, interlayer coupling causes Josephson couplings between adjacent layers enhancing 3D-like behavior (12), which is being increasingly affected by orbital pair breaking effect. As a result, Ising superconductivity although configured for each individual layer cannot manifest itself in bulk phases. To preserve the pinning in individual layers, field-induced superconducting devices, being able to electrically control the isolation of a single layer of superconducting $MoS_2$ (26), provide an unique way to realize Ising superconductor in TMD systems.

However, the reduced $\boldsymbol{B}_{\text{eff}}$ might still be a candidate mechanism for the large in-plane $B_{c2}$ found in bulk superconducting TMDs (32,37–40,54,55), where similar idea of being orthogonal to $\boldsymbol{B}_{\text{ex}}$ is implemented theoretically by out-of-plane components either originated form SOS (12) or lattice polarization (36). Experimentally, signatures of spin protection manifest as enhanced $B_{c2}$ in systems with lowering lattice symmetry by intercalating organic molecules and alkali elements with large radius (Cs doped $MoS_2$ shows highest $B_{c2}$ among bulk phases in Fig. 3D) or by forming charge density wave which polarized the in-plane lattice (54). The present result points out a unified physical picture for the increase of $B_{c2}$, where lowering symmetry could unveil the hidden Ising superconductivity in individual TMD layer.

Considering other well-known superconducting systems, where spins of Cooper pairs are configured in alternative ways against external magnetic field, we compare our $B_{c2}$ with those obtained from other superconductors under their maximum spin protection (along the labeled crystal axis). For 3D Rashba protected and spin triplet superconductors, although aligned spins of Cooper pairs can be free from $B_p$, nevertheless, $B_{c2}$ is still affected by orbital pair breaking effect in both cases. Whereas, in gate-induced Ising superconductor, in addition to strong pining of spins that overcomes the $B_p$, the orbital effect is also eliminated by the 2D nature of the system. Therefore, Ising superconductor exempt from both orbital effect and the paramagnetic effect can have $B_{c2}$ values well above $B_p$ (red dashed line in Fig. 4F) exhibiting a high $B_{c2}/B_p$ ratio. The Zeeman field protected states in Ising superconductor is clearly among the most robust superconducting states available against external magnetic field. Given the very similar band structures found in 2$H$-type TMDs with universal Zeeman-type spin splitting, we would expect a family of Ising superconductors in 2$H$-type TMDs because of the growing recent successes in inducing more TMD superconductors using field effect (21,56,57). The concept of Ising superconductor is also applicable to other layered systems, where similar intrinsic SOC could be induced by symmetry breaking.

**Figure 1** **Inducing superconductivity in MoS$_2$ thin flakes by gating.** (**A**) Conduction band electron pockets near the *K* and *K'* points the hexagonal Brillouin zone of monolayer MoS$_2$. Electrons in opposite K and K' points experience opposite effective magnetic field ***B***$_{eff}$ and -***B***$_{eff}$ denoted by the out-of-plane green arrows, respectively. The blue and red coloured pockets indicate electron spins orientating up and down, respectively. (**B**) The side view (left panel) and top view (right panel) of top four layers in a multi-layered MoS$_2$ flake. The vertical dashed lines show the relative positions of Mo and S atoms in 2*H*-type stacking. In-plane inversion symmetry is broken in each individual layer but global inversion symmetry is restored in bulk after stacking. (**C**) Energy band splitting caused by ***B***$_{eff}$, where blue or red band denote spin aligned up or down, respectively. Due to 2*H*-type stacking, adjacent layers have opposite ***B***$_{eff}$ at the same *K* points. (**D**) The red curve (left axis) denotes the carrier density $n_{2D}$ for the top four layers of MoS$_2$ for sample D1 with $T_c(0)$ = 2.37 K. When superconductivity in the first layer is induced near the onset of superconducting phase in the established MoS$_2$ phase diagram (right axis), the second adjacent layer is already below the onset of metallic transport (black dot) and the rest layers are all insulating. In the phase diagram (right axis), each superconducting state with different $T_c(0)$ is denoted by a fixed colour, which is consistently used throughout all other figures. Here, $T_c$ is determined at the temperature where resistance drop reaches 90% of its normal state resistivity $R_N$ at 15 K. This criteria is different from 50% $R_N$ used in the rest of the paper, which is intentionally chosen to be consistent with that used in phase diagram established previously (21). (**E**) Temperature dependence of sheet resistivity $R_s$ showing different $T_c$ corresponding to superconducting states (from sample D1 and D24) denoted in panel **D** in same colours.

**Figure 2** **Two-dimensional superconductivity in gated MoS$_2$ (Sample D24).** Panels (**A**) and (**B**) show the temperature dependence of sheet resistance $R_s$ under constant out-of- and in-plane magnetic field up to 12 T, respectively. In panel **B**, the left insert shows expanded details near half of the normal state resistivity $R_N/2$ within $\Delta T \approx 1$ K. The right insert illustrates the definition of $\theta$, which is the angle between the *B* field and the MoS$_2$ surface. (**C**) Angular dependence of $R_s$, where the dashed line denotes $R_s = R_N/2$. Inset: data expanded within $\Delta\theta \approx \pm 1°$ near in-plane field configuration ($\theta = 0°$). (**D**) Angular dependence of $B_{c2}$, which is fitted by both 2D Tinkham (red) and 3D anisotropic

Ginzburg-Landau model (blue). Inset: angular dependence of $B_{c2}$ expanded within $\Delta\theta \approx \pm 1°$ near in-plane field configuration ($\theta = 0°$). (**E**) *V–I* relationship at different temperatures close to $T_c$ plotted in logarithmic scale. The black lines are fitting close to metal–superconductor transitions. The long black line denotes $V \propto I^3$, which gives the $T_{BKT}$. (**F**) Temperature dependence of $\alpha$ from fitting the power law dependence of $V \propto I^\alpha$ of the black lines in panel **E**. $T_{BKT} = 6.3$ K is obtained for $\alpha = 3$.

**Figure 3** **Determining the in-plane upper critical field $B_{c2}$ at different $T_c$ (sample D1 and D24)**. (**A**) Magnetoresistance of sample D1 (with $T_c(0) = 2.37$ K near the onset of superconducting phase) as a function of in-plane magnetic field up to 37 T at various temperatures. (**B**) Temperature dependence of $R_s$ of sample D1 (with $T_c(0) = 7.64$ K) under different in-plane magnetic fields up to 32.5T. The dashed lines in panels **A** and **B** indicate half of the normal state resistivity $R_N/2$. Here, $B_{c2}$ is determined as the intercepts between dashed lines and $R_s$ curves. (**C**) Temperature dependence of $B_{c2}$ for superconducting states induced in sample D1 with different $T_c$ (filled circles). The $B_{c2}$ obtained from alkali metal intercalated bulk $MoS_2$ is extracted from Ref. (48) as comparison. The $B_{c2}$ is fitted as a function of temperature with 2D G–L (solid line) and KLB (dashed line) models. (**D**) Normalized $B_{c2}/B_p$ plotted as a function of reduced temperature $T/T_c$ including superconducting states from alkali doped bulk phases and gated-induced phases (sample D1 and D24). The dashed line denotes $B_p$ and sets boundary of Pauli limited regime (grey shaded).

**Figure 4** **Interplay between external magnetic field and spins of Cooper pairs aligned by Zeeman and Rashba-type effective magnetic fields**. (**A-D**) Schematic illustration of acquiring Zeeman energy through coupling between an external magnetic field and the spins of Cooper pairs formed near the *K* and *K'* points of the Brillouin zone (not in scale). For the cases shown in **A** and **C**, when Rashba and Zeeman SOC align the spins of Cooper pairs parallel to the external field, significant increase of Zeeman energy due to parallel coupling between field and spin can eventually cause pair breaking. In contrast, in **B** and **D**, the acquired Zeeman energy is minimized due to the orthogonal coupling between the field and aligned spins, which effectively protects the Cooper pairs from depairing. (**E**) Theoretical fitting of $B_{c2}/B_p$ versus $T/T_c$ relationship for sample D1 ($T_c(0) = 2.37$ K, 5.5 K) and D24 ($T_c(0) = 7.38$ K) using a fixed effective Zeeman field ($\beta_{so} = 6.2$ meV) and an increasing Rashba field ($\alpha_R k_F$ range from 10% to ~ 50% of $\beta_{so}$). Two dashed lines show the special cases when only Rashba field ($\alpha_R k_F$ = 30 meV, $\beta_{so}$ = 0) is considered (red) and both Zeeman and Rashba fields are zero (black). In the former case, a large $\alpha_R k_F$ causes a moderate increase of $B_{c2}$ to ~$\sqrt{2}B_p$ (10). In the latter, the conventional Pauli limit at zero temperature is recovered. (**F**) Plot of $B_{c2}$ versus $T_c$ for different superconductors (magnetic field was applied along a specific crystal axis such as *a*, *b* and *c* or to a polycrystalline denoted by *poly*). This collection contains well-known systems of non-centrosymmetric (pink circles), triplet (purple square) (6,8,9), low-dimensional organic superconductors (green triangle) (42,58–60) and bulk superconducting TMD systems (blue triangle) (37–40,55). Robustness of the spin protection can be measured by the vertical distance between $B_{c2}$ and the red dashed line denoting $B_p$. It is evident that gatie-induced superconductivity from samples D1 and D24 are among states showing the highest $B_{c2}/B_p$ ratio. In (LaSe)$_{1.14}$(NbSe$_2$), $T_c$ was determined at 95% of $R_N$, $T_c$ in organic molecule intercalated TMDs was obtained by extrapolating to zero resistance, all other systems use the standard of 50% of $R_N$.

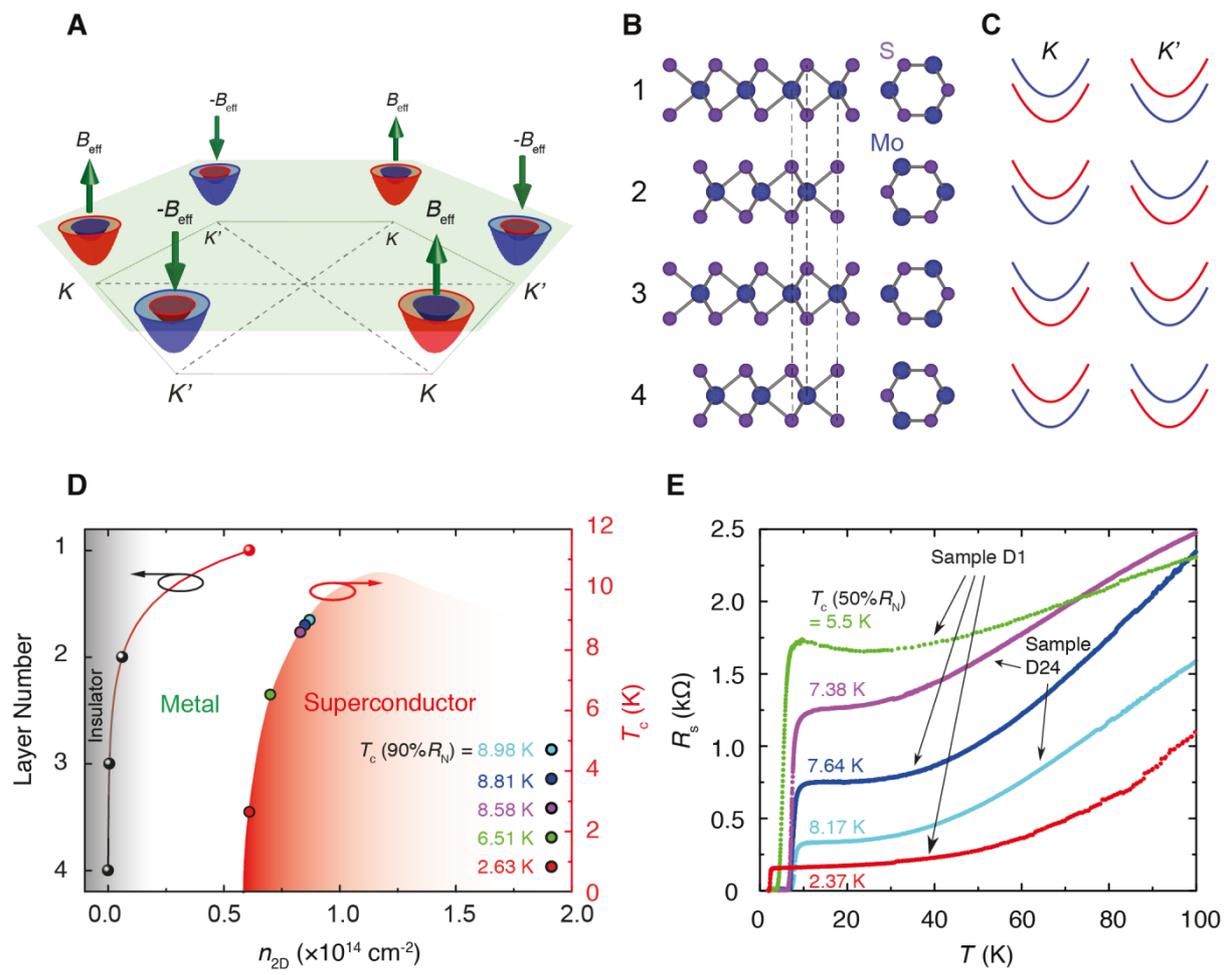

**Figure 1**

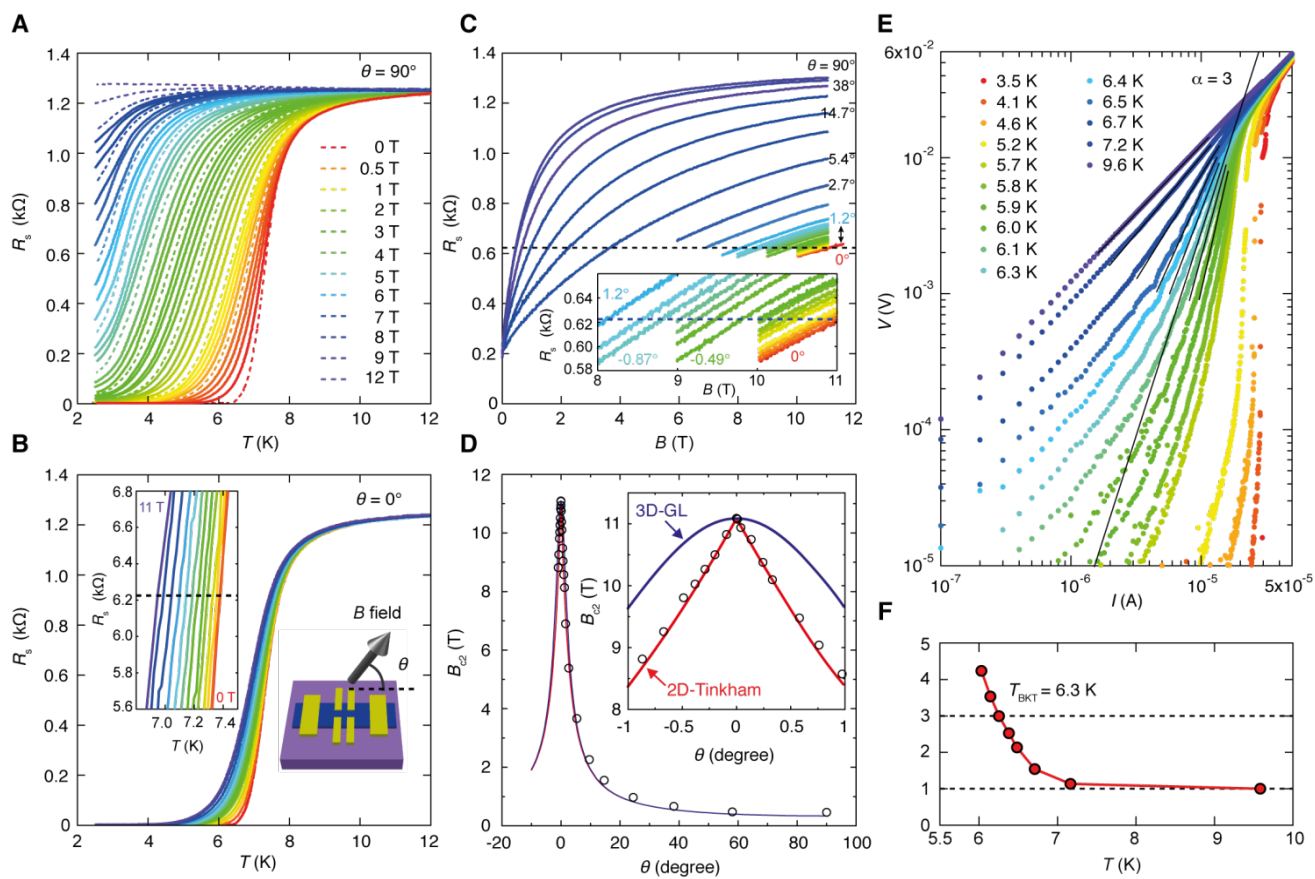

**Figure 2**

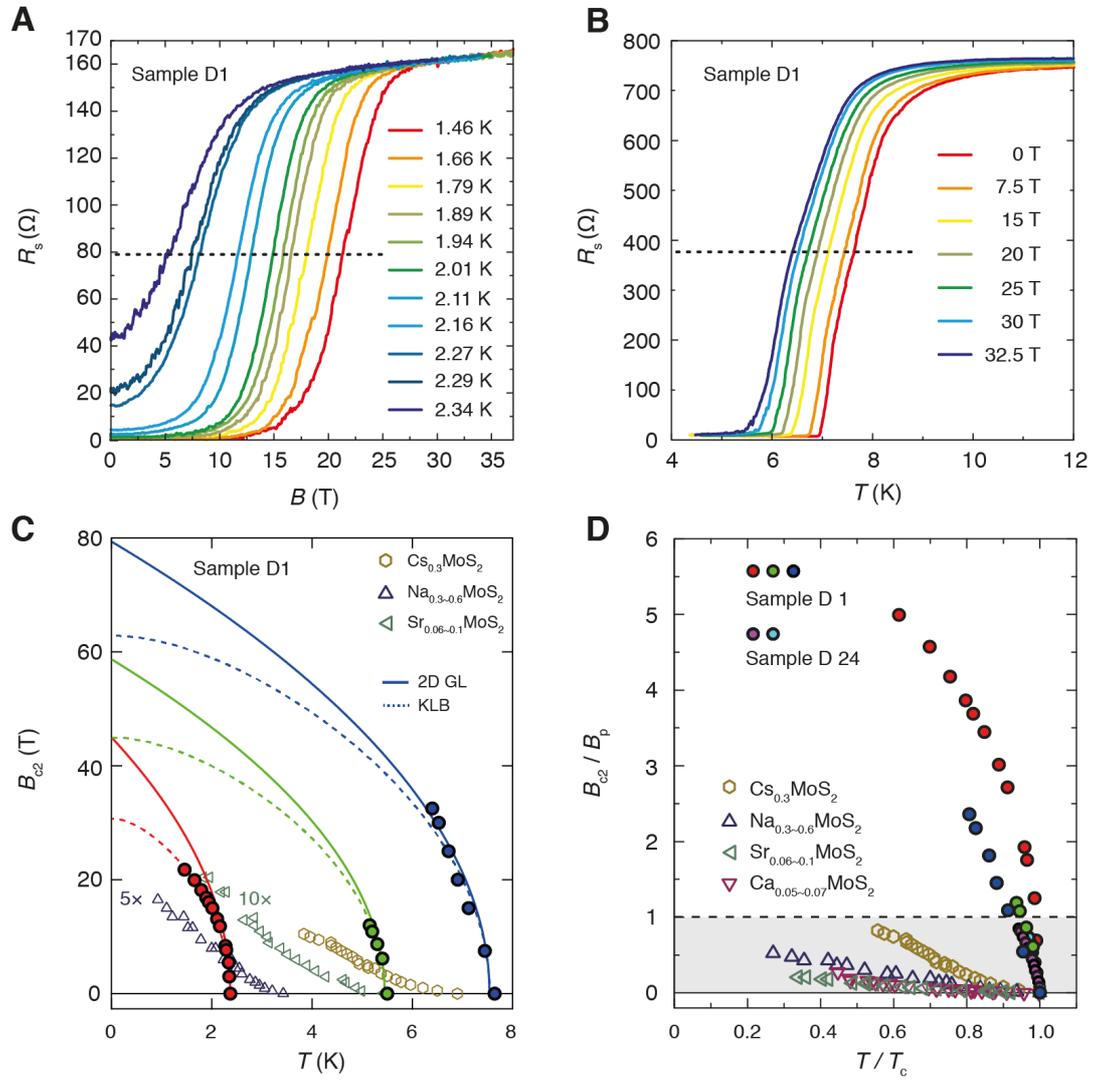

**Figure 3**

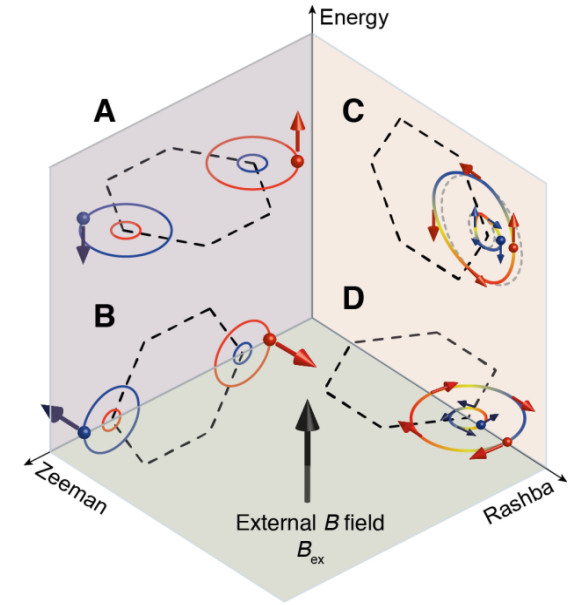
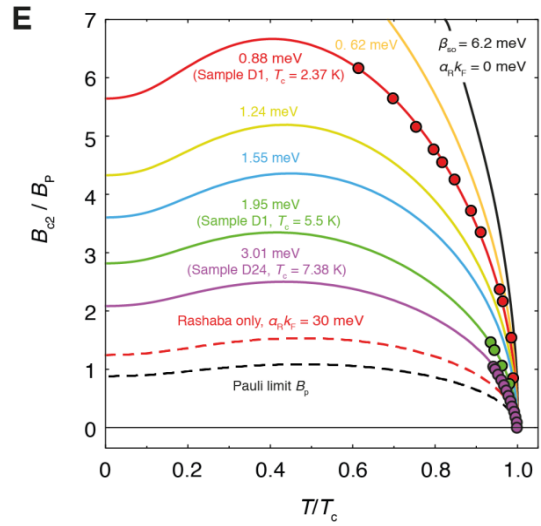
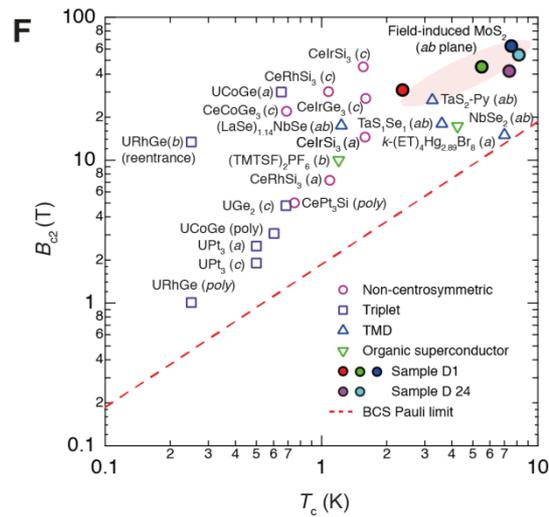

**Figure 4**